\documentclass[useAMS,usenatbib]{mn2e}
\usepackage{amssymb}
\usepackage{times}
\usepackage{epsfig}
\usepackage{longtable}
\usepackage{graphicx}

\newcommand{\ud}{\mathrm{d}}

\newcommand{\be}{\begin{equation}}
\newcommand{\ee}{\end{equation}}
\newcommand{\beq}{\begin{eqnarray}}
\newcommand{\eeq}{\end{eqnarray}}
\newcommand{\beqno}{\begin{eqnarray*}}
\newcommand{\eeqno}{\end{eqnarray*}}
\newcommand{\bitmz}{\begin{itemize}}
\newcommand{\eitmz}{\end{itemize}}

\def\Rs{R_{\rm S}}

\def\Rco{R_{\rm co}}
\def\Rm{R_{\rm m}}
\def\Rt{R_{\rm t}}

\def\Rdisc{R_{\rm disc}}

\def\Rbb{R_{\rm bb}}
\def\Rseed{R_{\rm seed}}

\def\Rns{R_*}
\def\Msun{{\rm M}_{\odot}}
\def\Mns{M_*}

\def\Mdot{\dot{M}}

\def\Lbol{L_{\rm bol}}

\def\Ldisc{L_{\rm disc}}

\def\Fbol{F_{\rm bol}}
\def\Fdisc{F_{\rm disc}}

\def\Teff{T_{\rm eff}}

\def\Tdisc{T_{\rm disc}}
\def\Kdisc{K_{\rm disc}}
\def\Tbb{T_{\rm bb}}
\def\Kbb{K_{\rm bb}}
\def\Tseed{T_{\rm seed}}
\def\Kseed{K_{\rm seed}}
\def\Te{T_{\rm e}}

\def\KFe{K_{\rm Fe}}

\def\rhoin{\rho_{\rm in}}
\def\rhoinb{\rho_{\rm in,1}}
\def\rhoina{\rho_{\rm in,2}}
\def\rhout{\rho_{\rm out}}

\def\NH{N_{\rm H}}

\def\refl{{\Re}}

\def\omegans{\omega_*}

\def\chired{\chi^2_{\rm red}}

\def\ergs{{\rm erg\,s^{-1}}}
\def\ergcm2s{{\rm erg\,cm^{-2}\,s^{-1}}}

\def\Bs{B_{\rm S}}

\def\Bz{B_{z}}
\def\Bphi{B_{\phi}}

\def\sax{SAX J1808.4--3658}
\def \xte {{\it RXTE}}
\def\swift{{\it Swift}}
\def\xmm{{\it XMM-Newton}}

\def\apj{ApJ}%
\def\apjl{ApJ}%
\def\aap{A\&A}%
\def\mnras{MNRAS}%
\def\pasj{PASJ}%
\def\apjs{ApJS}%
\def\nat{Nat}
\def\apss{Ap\&SS}
\title[Varying disc--magnetosphere coupling in \sax]{Varying disc--magnetosphere coupling as the origin of pulse profile variability in \sax}
\author[Jari J. E. Kajava et al.]{Jari\,J.\,E.\,Kajava,$^{1}$\thanks{E-mail: jari.kajava@oulu.fi, juri.poutanen@oulu.fi} Askar Ibragimov,$^{1,2}$ Marja Annala,$^1$ Alessandro Patruno$^3$  and Juri Poutanen$^1$\footnotemark[1] \\
$^1$Astronomy Division, Department of Physics, P.O.Box 3000, FI-90014 University of Oulu, Finland \\
$^2$Sabanc{\i} University,  Orhanl{\i}-Tuzla, Istanbul, 34956, Turkey \\
$^3$Astronomical Institute ``Anton Pannekoek," University of Amsterdam, Kruislaan 403, 1098 SJ Amsterdam, The Netherlands \\
}

\begin{document}

\date{Accepted 2011 June 30. Received 2011 June 30; in original form 2011 March 23}

\pagerange{\pageref{firstpage}--\pageref{lastpage}} \pubyear{2011}

\maketitle

\label{firstpage}

\begin{abstract}
Accreting millisecond pulsars show significant variability of their pulse profiles, especially at low accretion rates. 
On the other hand, their X-ray spectra are remarkably similar with not much variability over the course of the outbursts. 
For the first time, we have discovered that during the 2008 outburst of \sax\ a major pulse profile change was accompanied by a dramatic variation of the disc luminosity at almost constant total luminosity.
We argue that this phenomenon is related to a change in the coupling between the neutron star magnetic field and the accretion disc. 
The varying size of the pulsar magnetosphere can influence the accretion curtain geometry and affect the shape and the size of the hotspots.
Using this physical picture, we develop a self-consistent model that successfully describes simultaneously the pulse profile variation as well as the spectral transition. 
Our findings are particularly important for testing the theories of accretion onto magnetized neutron stars,  
better understanding of the accretion geometry as well as the physics of disc-magnetosphere coupling.   
The identification that varying hotspot size can lead to pulse profile changes has profound implications for determination of the neutron star masses and radii. 
\end{abstract}

\begin{keywords}
accretion, accretion discs -- pulsars: individual: SAX J1808.4--3658 -- stars: neutron -- X-rays:  binaries 
\end{keywords}

\section{Introduction}\label{intro}

\sax\ was first detected with the Wide Field Cameras on board the  {\it BeppoSAX} satellite in 1996 \citep{itz98}. 
In 1998, another outburst of \sax\ was observed with {\it Rossi X-ray Timing Explorer} (\xte) and the discovery of $\approx 401$ Hz pulsations led to its identification as the first accreting millisecond pulsar \citep[AMP;][]{WvdK98, CM98}.
Since 1998, the source has been in outburst roughly every $2.5$ years (2000, 2002, 2005 and 2008). 
During the outbursts of \sax\ (and other AMPs), the magnetic field of the neutron star (NS) channels the accreted matter on to the stellar magnetic poles.
The emitted radiation from these hotspots is then modulated at the stellar spin period, that results in coherent pulsations.
In a typical outburst of \sax\ \citep[see fig.2 in][]{HPC08} the source rises from quiescence in roughly five days, reaching a peak flux of about $2\times10^{-9}\, \ergcm2s$ in the 2--25 keV range of the \xte\ Proportional Counter Array (PCA). 
After the peak, the source flux drops slowly (``slow decay" stage) in the course of the next ten days, although the 2008 outburst had a double peaked light curve \citep{HPC09}. 
The slow decay stage is followed by the ``rapid drop" stage, where the flux drops by an order of magnitude in a couple of days after which the source exhibits several re-brightening episodes every five days or so (``flaring tail" stage, see also \citealt{PW09}, \citealt{IP09}).

The system parameters of \sax\ are the best known among AMPs. 
The orbital period is $\approx 2$ hours \citep{CM98, BRD09, HPC09} and the companion star has a very low mass of $M_{\rm c} \sim 0.05 \Msun$ \citep{BC01}.
The analysis of type I X-ray bursts by \citet{GC06} gave a distance estimate of $3.5 \pm 0.1$ kpc.
The inclination has been constrained to $i=$36$^\circ$--67$^\circ$ using optical observations \citep{DHT08} and the X-ray analysis \citep{IP09} of the 2002 outburst led to a similar constraint of $i \approx 50^\circ$--$70^\circ$.
By studying the long term timing evolution of the system, \citet{HPC08} constrained the neutron star surface magnetic dipole field to a range of $\Bs = (0.4$--$1.5) \times 10^8\,{\rm G}$. 
Another estimate of $ \Bs = (0.8 \pm 0.5) \times 10^8 k_{\rm A}^{-7/4}\,{\rm G}$ from pulse profile variability, was derived by \citet{IP09}, where the factor $k_{\rm A} \approx 0.5$ \citep{long05}.
Also, \citet{BDM06} gave an estimate of $ \Bs \sim (3.5 \pm 0.5) \times 10^8\,{\rm G}$ and \citet{PWvdK09} obtained $\Bs = (2$--$3) \times 10^8\,{\rm G}$.

The energy spectra of \sax\ has been studied extensively \citep[e.g.][]{GR98,GDB02,PG03,IP09}. 
It was pointed out already by \citet{GR98} that the spectral shape remains remarkably similar when flux changes by 
more than an order of magnitude throughout the outburst.  
The energy spectrum is roughly flat (photon index $\Gamma \approx 2$) with a cutoff/roll-over at energies above $\sim 50$ keV. 
Such spectra -- also seen in other AMPs \citep{P06} -- are most likely produced by thermal Comptonization of soft seed photons originating from the heated stellar surface by hot electrons in the accretion shock \citep{GDB02,PG03,IP09}.   
Below $\sim 5$ keV an additional $T \approx 0.5$ keV thermal component is visible that is associated with the heated surface of the NS. 
\xmm\ observations have revealed another {\it non-pulsating} cooler thermal component due to the accretion disc \citep{PRA09}.
The X-ray emission from the NS surface irradiates the accretion disc, that causes spectral hardening above 10 keV due to Compton reflection \citep{IP09}.
This irradiation also produces an iron line at 6.4 keV and the modelling of the line profiles from \xmm\ and {\it Suzaku} observations by \citet{PDD09} and \citet{CAP09} has given tight constraints on inclination and the accretion disc truncation radius. We note, however, that these results depend strongly on the continuum spectral model and data reduction issues (see \citealt{NDC10}, for discussion).

Although many aspects of AMP physics are known, one of the unresolved puzzles is the origin of sudden pulse profile changes in \sax\ \citep{HPC08,HPC09} and other AMPs.
There are many mechanisms that can cause such changes \citep{P08} and they contribute to the ``timing noise" seen in many AMPs \citep[e.g.][]{BDM06,pap07,RDB08,PWvdK09, PIA09}.
One of these remarkable pulse profile changes occurred in the 2008 outburst of \sax\ \citep{HPC09}.
In the beginning of the slow decay stage, the pulse profile was rather symmetric showing only small harmonic content.
However, on September 27 the fundamental pulse amplitude decreased by almost 50 per cent, while no change was seen in other observed quantities, especially the observed flux and phases of the harmonics.
A few days later on October 2 (MJD 54742), the pulse profile quickly morphed to have two peaks. 
This time the pulse profile transition was accompanied by a decay in the observed flux and jumps in pulse phases. 
The double peaked profile was seen only for about three days, because on October 6 (MJD 54746), right before the onset of the rapid drop stage, the fundamental amplitude increased significantly and the profile changed again to a shape similar to what was seen before September 27.
This behaviour cannot be explained by a change in the accretion disc truncation radius that was suggested by \citet{IP09} and \citet{PIA09}.
Although this model explains the jump in the fundamental phase (and the pulse profile change), that is associated with the rapid drop stage of the 2002 outburst of \sax\ \citep{BDM06}, it cannot be the case here simply because the observed flux (which can be related to the truncation radius) in these sudden pulse profile changes is almost an order of magnitude higher and it remains nearly constant \citep[see fig. 1 in][]{HPC09}. 
Among the 13 known AMPs \citep[see][and references therein]{Pat10}, some sources show similar jumps in the fundamental pulse amplitudes as in the 2008 outburst of \sax. 
A few examples include XTE J0929--314 \citep{GC02}, the first eclipsing AMP SWIFT J1749.4--2807 \citep{MS10,ACP11} and SWIFT J1756.9--2508 \citep{PAM10}.
Also, similar jumps in the fundamental pulse amplitude was seen in the 2002 and 2005 outbursts of \sax\ (see fig. 6 in \citealt{IP09} and fig. 1 in \citealt{HPC08}).

The variability in pulse amplitudes and profiles are sometimes accompanied with jumps in pulse phases \citep{pap07,RDB08,PWvdK09}.
In most cases, this timing noise is related with changes in the observed flux \citep{BDM06, IP09, PWvdK09} and the most notable changes usually occur in the fundamental phase \citep{BDM06,pap07,RDB08,PWvdK09}.
This type of timing noise is most likely caused by mass accretion rate dependent change of the hotspot location on the NS surface \citep{LBW09}, which explains the overall trends with flux and pulse phase residuals in \sax\ \citep{HPC08} and in many other AMPs \citep{PWvdK09}. 
However, although the general trends can be explained \citep{PWvdK09}, there are several AMPs where additional timing noise is clearly present. Most notable cases are the 2002, 2005 and 2008 outbursts of \sax, XTE J1807--294 and XTE J0929--314 (see fig. 3 in \citealt{PWvdK09} and fig. 1 in \citealt{HPC09}). In these cases another mechanism -- that is not related to changes in mass accretion rate -- must be responsible for the timing noise.

In this paper, we present our spectral and timing analysis of the 2008 outburst of \sax\ using \swift\ and \xte\ data. 
We find evidence that the sudden changes in the pulse profiles and in pulse amplitudes of \sax\ are driven by a changing interaction between the NS magnetic field and the accretion disc.

\section{Observations and data reduction}\label{observations}

The 2008 outburst of \sax\ was first detected on September 21 \citep{2008ATel.1728....1M}.
The outburst was very similar to the 2005 outburst in duration and brightness, albeit the flaring stage was dimmer \citep{HPC09}.
The slow decay stage of the outburst outburst was monitored by both \swift\ and \xte\ from September 24 to October 3, during which the fundamental pulse amplitude and the pulse profile changed significantly \citep[see][fig. 1]{HPC09}.
In this paper we perform a detailed spectral and pulse profile analysis of these data in order to investigate the origin of this sudden timing transition.

The \textit{RXTE} data (ObsID 93027) were reduced using {\sc heasoft} v.6.8 and the CALDB. 
We used data taken both by \textit{RXTE}/PCA (3--25 keV) and HEXTE (25--100 keV).
In cases where HEXTE exposures were short, we ignored the noisy channels above $\sim 60$ keV. 
Standard 0.5 per cent systematic was applied to the PCA spectra \citep{JMR06}. 
To keep the calibration uniform, we used data from PCU 2 only (all layers).
The \swift/XRT data (0.6--7 keV) were reduced using the {\sc xrtpipeline} v.0.12.3 in {\sc heasoft} v.6.8. 
The observations were performed in window-timing mode.
We used standard filtering and screening criteria for the event selection. 
Exposure maps were generated with the {\sc xrtexpomap} task and the ancillary response files with {\sc xrtmkarf} to account for different extraction regions (we used circular regions of 20 pixel radius), vignetting and psf corrections. The redistribution matrices (v.011) were taken from the CALDB.
The XRT spectra were then grouped using {\sc grppha} to have at least 20 counts in each bin.

We used the \swift/XRT data to model the time averaged spectra together with the quasi-simultaneous \xte\ pointings. 
There were a few cases where we could not use all the XRT data. 
The first XRT observation (observation 2, see Table \ref{tab:obslog}) was triggered by an X-ray burst.
We found that the XRT spectrum differed from the PCA spectrum taken before the X-ray burst and also from the XRT spectrum taken only 3 hours after the X-ray burst (observation 3). 
The spectra of observations 1 and 3 matched well indicating that the first XRT spectra was affected by the X-ray burst and it was therefore excluded from the analysis. 
Also, we did not use the observation 18, because the source was right on top of the bad columns.
In two occasions, where XRT observations were split into two snapshots (14 and 24), we used the data from the longer snapshot.

\begin{table} 
  \caption{The observation log.}
 %\centering
  \label{tab:obslog}
  \begin{tabular}{@{}rlcc}
  \hline
   \hline
\# &   ObsID  		& Date and time	 	& Exposure$^{a}$	 	\\
\hline
1 & 93027-01-01-07$^b$	& 2008-09-24 19:33:52--20:16:32 & $2333/763$\\
2 & 00325827000$^c$		& 2008-09-24 21:28:45--21:56:00	& $...$\\
3 & 00030034026			& 2008-09-24 23:17:40--23:32:00	& $816$\\
4 & 93027-01-01-06		& 2008-09-25 05:24:48--06:24:32 & $1035/414$\\
5 & 93027-01-01-05		& 2008-09-25 08:08:32--10:44:00	& $6482/2359$\\
6 & 00030034027			& 2008-09-25 13:34:28--13:52:00	& $1005$\\
7 & 93027-01-01-10		& 2008-09-25 19:52:16--20:36:32	& $490/123$\\
8 & 00030034028			& 2008-09-26 11:59:49--12:19:00	& $1142$\\
9 & 93027-01-02-00		& 2008-09-26 13:55:28--18:06:40 & $9713/3264$\\
10 & 93027-01-02-01		& 2008-09-26 23:32:16--23:50:40	& $1060/356$\\
11 & 00030034029			& 2008-09-27 04:04:15--04:23:00	& $1116$\\
12 & 93027-01-02-05		& 2008-09-27 10:24:00--15:00:32	& $8942/3411$\\
13 & 93027-01-02-06		& 2008-09-28 05:24:32--05:57:52 & $1943/668$\\
14 & 00030034030$^d$		& 2008-09-28 12:11:35--15:44:00	& $880$\\
15 & 93027-01-02-03		& 2008-09-28 16:13:52--17:35:44	& $3201/1043$\\
16 & 93027-01-02-04		& 2008-09-29 06:22:08--07:45:36	& $2775/1022$\\
17 & 00030034031			& 2008-09-29 10:44:25--10:59:00	& $849$\\
18 & 00030034032$^e$		& 2008-09-30 01:12:05--01:33:00	& $...$\\
19 & 93027-01-02-07		& 2008-09-30 09:03:12--10:01:36	& $3181/1154$\\
20 & 93027-01-02-09		& 2008-09-30 14:01:04--15:03:44	& $2306/744$\\
21 & 93027-01-02-02		& 2008-10-01 11:44:48--14:59:12	& $6637/2242$\\
22 & 00030034033			& 2008-10-01 14:25:42--14:45:00	& $1131$\\
23 & 93027-01-02-08		& 2008-10-02 14:25:52--17:19:44	& $6280/2035$\\
24 & 00030034034$^f$		& 2008-10-02 22:23:50--23:59:59	& $848$	\\
25 & 93027-01-03-00		& 2008-10-03 09:49:20--13:22:40	& $7557/2510$\\
\hline
%\multicolumn{4}{l}{$^{a}$XRT or PCA/HEXTE exposure times.}\\
%\multicolumn{4}{l}{$^{b}$Only data before X-ray burst used.}\\
%\multicolumn{4}{l}{$^{c}$Not used due to proximity to the X-ray burst.}\\
%\multicolumn{4}{l}{$^{d}$Only data from snapshot 2 used.}\\
%\multicolumn{4}{l}{$^{e}$Not used because the source was right on top of bad columns.}\\
%\multicolumn{4}{l}{$^f$Only data from snapshot 1 used.}
\end{tabular}
$^{a}$XRT or PCA/HEXTE exposure times (in seconds).
$^{b}$Only data before X-ray burst used.
$^{c}$Not used due to proximity to the X-ray burst.
$^{d}$Only data from snapshot 2 used.
$^{e}$Not used because the source was right on top of bad columns.
$^f$Only data from snapshot 1 used.
\end{table}

\begin{figure*}
\centerline{\epsfig{file= 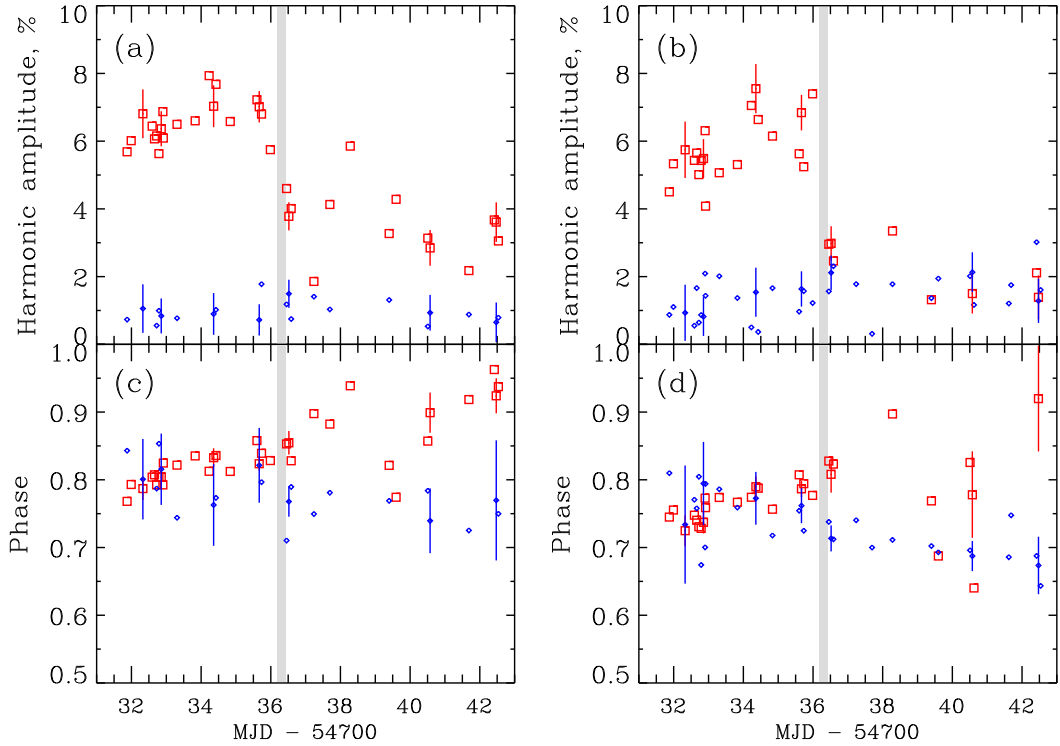,width=15cm}}
\caption{Change of the pulse profile parameters during the outburst. 
 Left panels are for the soft (3.7--5.7 keV) band, while the right panels are for  the hard (9.8--23.2 keV) band. 
 Upper panels present amplitudes of the fundamental  (red squares)
 and of the first overtone (blue diamonds). 
 Lower panels correspond to the pulse phases shown by the same symbols. 
Only those phases and amplitudes that can be constrained are plotted, and errors for only a few representative data points are plotted for visual clarity.
The time of the spectral transition is marked with the grey shaded strip (see Fig. \ref{fig:ratio_spectra}). Note the absence of a pulse phase shift during the transition (MJD 54736).}
\label{fig:ampl_phases}
\end{figure*}

\section{Spectral and timing analysis}\label{spectra}

\subsection{Simultaneous timing- and spectral transition}
\label{sec:profiles}

During the slow decay stage of the 2008 outburst -- on September 27 (MJD 54736) -- the  pulse amplitude dropped and the pulse profile changed significantly \citep[see][fig. 1]{HPC09}. 
The origin of these type of timing changes is not well understood, but they must be caused by some changes in the accretion geometry.
In order to analyse this timing transition, we construct pulse profiles at various energies and different time intervals using ephemeris from \citet{HPC09}. The observed pulse profiles are then fitted by a sum of two harmonics:
\be 
\label{harmonics}
F(\phi) = \bar{F} \{ 1 + a_1 \cos[2\pi (\phi - \phi_1)] + a_2 \cos[4\pi(\phi - \phi_2)] \},
\ee
where $F(\phi)$ is the flux at phase $\phi$, $\bar{F}$ is the mean flux and $a_1$, $a_2$, $\phi_1$ and $\phi_2$ are the amplitudes and phases of the fundamental and the first overtone, respectively.
The best-fitting amplitudes and phases are presented in Fig. \ref{fig:ampl_phases} for the soft (3.7--5.7 keV) and hard (9.8--23.2 keV) energy bands.
We selected these two bands because in the hard band we have emission only from the Comptonized component, whereas in the soft band there is a contribution from the blackbody component (e.g. \citealt{IP09}).
We see a drop in the fundamental amplitude and a change in the pulse profile on September 27 (MJD 54736), but we do not detect significant jumps in the pulse phases during the transition.

\begin{figure}
\centerline{\epsfig{file=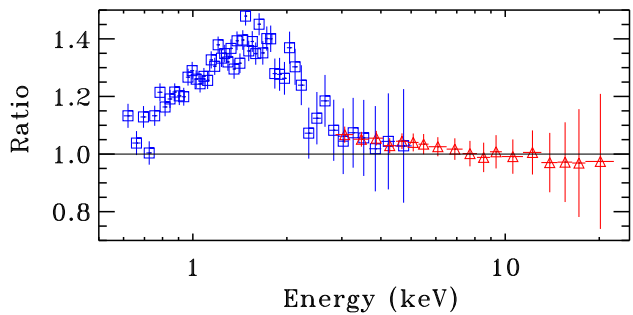}}
\caption{Ratio of the spectra from observations after and before the timing transition. The corresponding XRT spectra are from observations 14 and 11 (blue squares) and the PCA spectra are from observations 12 and 10 (red triangles). 
The spectral data are binned for visual clarity. The spectral transition occurred between 2008-09-27 04:23:00 and 2008-09-27 10:24:00 (see Table \ref{tab:obslog}). The increase of the flux is most evident below $\sim3$ keV, where the accretion disc component dominates (see \citealt{PRA09} and Fig. \ref{fig:illustr}).}
\label{fig:ratio_spectra}
\end{figure}

However, we find that the timing transition was accompanied with a simultaneous softening of the energy spectrum below $\sim5$ keV. 
This can be seen by taking a ratio of the observed spectra before and after the timing transition (from XRT observations 11 and 14 and PCA observations 10 and 12).
The ratio spectra are shown in Fig.~\ref{fig:ratio_spectra} and the softening is clearly seen in the \swift\ as well as in the \xte/PCA data.
The fact that the timing- and spectral transition occur simultaneously suggests a common physical origin, so a detailed broad band spectral analysis and pulse profile modelling of this transition is warranted.

\subsection{Spectral model} \label{modelling}

We modelled the spectra using {\sc xspec 12} \citep{Arn96}.
Errors are quoted at the 90\% confidence level and the errors in the fluxes were computed with the {\sc cflux} model in {\sc xspec}.
The reported luminosities are bolometric (calculated in the range of 0.01--500 keV from the best-fitting model) assuming a distance of $D=3.5$ kpc \citep{GC06}.

Our spectral model consist of {\sc constant $\times$ wabs $\times$ (diskbb $+$ bbodyrad $+$ diskline $+$ compps)}.
The spectral model is rather complex and has a large number of parameters.
The {\sc constant} model component accounts for the different instrument normalizations between \swift/XRT, \xte/PCA and HEXTE.
We fixed the normalization of PCA and allowed XRT and HEXTE normalizations to vary. 
For XRT, the normalizations varied in a tight range around $0.9$ between the different exposures, whereas HEXTE normalizations varied in the range of $0.52 - 0.62$.
The effect of interstellar absorption was taken into account using the {\sc wabs} model \citep{MM83}, which is parametrized by the absorption column $\NH$.
We modelled the accretion disc component with the {\sc diskbb} model \citep{M84}, which has two parameters; the inner disc temperature $\Tdisc$ and normalization (proportional to inner disc radius $\Rdisc$) $\Kdisc = [(\Rdisc/\mathrm{km})/(D/10\ \mathrm{kpc})]^2 \cos i$.
For the thermal emission from the NS surface, we used the {\sc bbodyrad} model, which is characterized by the black body temperature $\Tbb$ and normalization (proportional to the surface area) $\Kbb = [(\Rbb/\mathrm{km})/(D/10\ \mathrm{kpc})]^2$, where $\Rbb$ is the apparent black body radius as observed from infinity.
Model {\sc compps} \citep{PS96} -- characterizing Comptonization in the accretion column (we assumed slab geometry) -- is described by the following parameters: 
Thomson optical depth $\tau$ across the slab, electron temperature $\Te$, 
seed photon temperature $\Tseed$ for Comptonization,  and 
normalization (proportional to the surface area) $\Kseed = [(\Rseed/\mathrm{km})/(D/10\ \mathrm{kpc})]^2$, with $\Rseed$ corresponding to the size of the accretion column.
The Compton reflection component (included in {\sc compps}) is defined by the amplitude $\refl=\Omega/2\pi$ (where $\Omega$ is the solid angle covered by the cold reflector as viewed from the X-ray emitting hotspots), index $\alpha$ of the radial profile of the disc emissivity ($\propto r^{\alpha}$), inner- and outer disc radius, inclination $i$ and the iron line energy and normalization $\KFe$ ({\sc diskline}, \citealt{FRS89}). 
We note that the {\sc diskline} model inner disc radius can be substantially different from $\Rdisc$, 
which we derive from the {\sc diskbb} model normalization. 
The former is related to the disc truncation radius $\Rt$, whereas the latter traces the radius where the dissipation ceases, which is related to the magnetospheric radius $\Rm$ (see end of Section \ref{sec:spectra}).

When we attempted to fit all these parameters to the data, we found several unphysical changes in the best-fitting parameters throughout the outburst.
Most notable inconsistencies were variable inclination $i$ and a larger hotspot radius than a typical NS radius.
Furthermore, several parameters showed tight correlated changes among them.
$\Te$, $\tau$ and $\refl$ -- which together define the spectral slope above $\sim$10 keV -- were strongly correlated, resulting in large uncertainties in these parameters. 
Also, $\NH$ showed inconsistent changes during the outburst.
We thus had to fix several parameters to obtain reasonable values.
We did this in an iterative manner, where we fixed certain parameters, then fitted the data again, inspected the results and then fixed more parameters if necessary.

Initially we fixed five parameters.
Inclination was fixed to $i = 60\degr$ (for both reflection and iron line, as in \citealt{IP09}). 
The inner disc radius that affects both the smearing of the line and Compton reflection was fixed to a value of $10 \Rs$, where $\Rs = 2GM/c^2$ is the Schwarzschild radius. The outer disc radius was fixed to 1000$\Rs$.
The radial disc emissivity for reflection was fixed to $\alpha = -3$.
Also, because of poor energy resolution of PCA and poor statistics of XRT spectra in the iron line region, we fixed the iron line energy to 6.4 keV.

The model as such described the data very well (the best-fitting parameters and fitting statistics were compatible to the ``final results" presented in Table \ref{tab:bestfits}). 
However, as discussed in \citet{GP05} and \citet{IP09}, the hotspot surface area ($\propto \Kbb$) was poorly constrained and in many cases resulted in large and unphysical values.
Furthermore, some residuals were present in the XRT data around 2 keV, where the calibration is known to have problems because of the Si K$\alpha$ edge at 1.839 keV \citep{GBA09}.
These residuals created a problem in the fitting, because the flux contribution of the $\Tbb \approx$0.5--0.6 keV NS blackbody component and the  $\Tdisc \approx$0.3--0.4 keV disc component was very similar at this energy.
We identified two cases (6 and 11) -- where the XRT observations were about $3\arcmin$ off-axis -- in which these instrumental residuals were the main reason why $\Tdisc$, $\Kdisc$, $\Tbb$, $\Kbb$ and $\NH$ seemed to change in a correlated way during the outburst.
The best-fitting value of $\NH$ varied roughly in a range of (0.9--0.95)$\pm$0.15$\times10^{21}\,{\rm cm}^{-2}$ for most of the observations, except in these two cases where it dropped to a value of $\sim(0.7 \pm 0.05) \times 10^{21}\,{\rm cm}^{-2}$.
As we do not expect $\NH$ to vary and its value was tightly correlated with the parameters of the accretion disc ($\Tdisc$ and $\Kdisc$), we fixed this parameter to the Galactic value of $1.13 \times 10^{21}\,{\rm cm}^{-2}$ (obtained from the {\sc headas} tool {\sc nh}) to avoid spurious changes in the disc parameters.
Also, following \citet{IP09}, we set $\Kbb = \Kseed$ in the fitting, which helped to remove the correlation between $\Tbb$ and $\Tdisc$ described above. 

After another round of spectral fitting, we found that $\Te \approx$(45--55)$\pm$10 keV in all the spectral fits.
In some cases the spectra were very noisy above $\sim$60 keV and some small changes in $\Te$ and $\tau$ between different observations were identified to be caused by one or two spectral channels only.
As we did not see any statistically important trend between the observations, we fixed $\Te$= 50 keV, which was the best-fitting constant value.
We then fitted the data again, and got physically acceptable values for all the model parameters. 
We also checked, that the removal of XRT data in the 1.6--2.4 keV range did not change the best-fitting parameters significantly.

\begin{table*}
 \begin{minipage}{\linewidth}
  %\centering
% \footnotesize 
  \caption{Best-fitting parameters. The fixed parameters were: $\NH = 1.13 \times 10^{21}\,{\rm cm}^{-2}$, $\Te = 50$ keV, $i=60\degr$. }
  \label{tab:bestfits}
 \begin{tabular}{r c c c c c c c c c c l}
\hline \hline
\#$^a $ &  $\Tdisc$ & $\Kdisc$ 			& $\Tbb$ & $\KFe$ 				& $\Tseed$ & $\tau$ & $\refl$ & $\Kbb$=$\Kseed$ 		& $\Fdisc$$^b$ & $\Fbol$$^b$ & $\chired/$d.o.f. \\
 &   (keV) & (km/10 kpc)$^2$  & (keV)   & $[\times 10^{-3}]$   & (keV)      &        &         & (km/10 kpc)$^2$ & 		   &  	 	 & 					\\
  \hline 
%\multicolumn{12}{c}{00030034026 \& 93027-01-01-07} \\
1, 3 & $ 0.28_{-0.02}^{+0.02}$ & $ 1250_{-270}^{+380}$ & $ 0.53_{-0.10}^{+0.07}$ & $  1.50_{-0.35}^{+0.36}$ & $ 0.86_{-0.09}^{+0.07}$ & $ 1.05_{-0.08}^{+0.11}$ & $ <0.40$ & $ 380_{-110}^{+220}$ & $ 1.60_{-0.07}^{+0.08}$ & $ 5.15_{-0.14}^{+0.17}$ &  1.07/  371 \\
%\multicolumn{12}{c}{00030034027 \& 93027-01-01-05} \\
5, 6 & $ 0.29_{-0.01}^{+0.01}$ & $ 1070_{-140}^{+170}$ & $ 0.51_{-0.06}^{+0.06}$ & $  1.50_{-0.25}^{+0.25}$ & $ 0.84_{-0.06}^{+0.06}$ & $ 1.05_{-0.05}^{+0.06}$ & $ 0.21_{-0.12}^{+0.11}$ & $ 430_{-100}^{+150}$ & $ 1.70_{-0.05}^{+0.05}$ & $ 5.37_{-0.09}^{+0.10}$ &  1.21/  550$^c$ \\
%\multicolumn{11}{c}{00030034028 \& 93027-01-02-00} \\
8, 9 & $ 0.28_{-0.01}^{+0.01}$ & $ 1130_{-150}^{+220}$ & $ 0.52_{-0.05}^{+0.04}$ & $  1.55_{-0.22}^{+0.23}$ & $ 0.84_{-0.05}^{+0.04}$ & $ 1.05_{-0.05}^{+0.05}$ & $ 0.21_{-0.10}^{+0.11}$ & $ 410_{-70}^{+130}$ & $ 1.47_{-0.04}^{+0.04}$ & $ 5.04_{-0.08}^{+0.08}$ &  1.07/  545 \\
%\multicolumn{11}{c}{00030034029 \& 93027-01-02-01} \\
10, 11 & $ 0.29_{-0.02}^{+0.01}$ & $ 1040_{-160}^{+230}$ & $ 0.52_{-0.12}^{+0.07}$ & $  1.62_{-0.43}^{+0.44}$ & $ 0.82_{-0.11}^{+0.07}$ & $ 1.09_{-0.11}^{+0.08}$ & $ <0.34$ & $ 450_{-120}^{+350}$ & $ 1.69_{-0.06}^{+0.05}$ & $ 5.29_{-0.17}^{+0.15}$ &  1.22/  553$^c$ \\
%\multicolumn{11}{c}{00030034030 \& 93027-01-02-06} \\
13, 14 & $ 0.40_{-0.02}^{+0.01}$ & $  400_{-50}^{+60}$ & $ 0.64_{-0.09}^{+0.09}$ & $  1.51_{-0.45}^{+0.46}$ & $ 1.00_{-0.09}^{+0.09}$ & $ 1.05_{-0.08}^{+0.08}$ & $ <0.42$ & $ 200_{-60}^{+100}$ & $ 2.17_{-0.08}^{+0.07}$ & $ 5.53_{-0.14}^{+0.25}$ &  1.01/  535 \\
%\multicolumn{11}{c}{00030034031 \& 93027-01-02-04} \\
16, 17 & $ 0.42_{-0.02}^{+0.02}$ & $  320_{-40}^{+50}$ & $ 0.66_{-0.08}^{+0.07}$ & $  1.59_{-0.40}^{+0.40}$ & $ 1.00_{-0.09}^{+0.08}$ & $ 1.09_{-0.09}^{+0.12}$ & $ <0.35$ & $ 180_{-50}^{+80}$ & $ 2.17_{-0.08}^{+0.08}$ & $ 5.38_{-0.15}^{+0.19}$ &  1.08/  533 \\
%\multicolumn{11}{c}{00030034033 \& 93027-01-02-02} \\
21, 22 & $ 0.33_{-0.01}^{+0.01}$ & $  650_{-80}^{+90}$ & $ 0.55_{-0.06}^{+0.06}$ & $  1.44_{-0.22}^{+0.22}$ & $ 0.87_{-0.06}^{+0.06}$ & $ 1.02_{-0.06}^{+0.08}$ & $ 0.27_{-0.16}^{+0.16}$ & $ 280_{-70}^{+110}$ & $ 1.68_{-0.05}^{+0.05}$ & $ 4.42_{-0.08}^{+0.10}$ &  1.11/  533 \\
%\multicolumn{11}{c}{00030034034 \& 93027-01-03-00} \\
24, 25 & $ 0.32_{-0.01}^{+0.01}$ & $  590_{-90}^{+110}$ & $ 0.52_{-0.07}^{+0.06}$ & $  1.30_{-0.18}^{+0.18}$ & $ 0.80_{-0.07}^{+0.06}$ & $ 0.98_{-0.08}^{+0.10}$ & $ 0.38_{-0.21}^{+0.23}$ & $ 300_{-80}^{+130}$ & $ 1.29_{-0.04}^{+0.04}$ & $ 3.52_{-0.07}^{+0.09}$ &  1.02/  484 \\
\hline
%\multicolumn{12}{l}{$^a$ Observation numbers from Table \ref{tab:obslog}.} \\
%\multicolumn{12}{l}{$^{b}$ Poor fits are mainly due to large residuals in XRT spectra around $\approx 2$ keV, which are most likely due to a calibration issue at the Si K$\alpha$ edge at 1.839 keV \citep{GBA09}. Removing the data between 1.6--2.4 keV improves the fits to $\chired \approx 1.1$, without a significant 
%change in the best-fitting parameters.} \\
\end{tabular}
$^a$ Observation numbers from Table \ref{tab:obslog}. 
$^b$ Flux in units of $10^{-9}\,\ergcm2s$.
$^c$ Poor fits are mainly due to large residuals in XRT spectra around $\approx 2$ keV, which are most likely due to a calibration issue at the Si K$\alpha$ edge at 1.839 keV \citep{GBA09}. Removing the data between 1.6--2.4 keV improves the fits to $\chired \approx 1.1$, without a significant change in the best-fitting parameters.
\end{minipage}
\end{table*}

\subsection{Results of the spectral modelling} 
\label{sec:spectra}

\begin{figure}
\includegraphics[width=84mm]{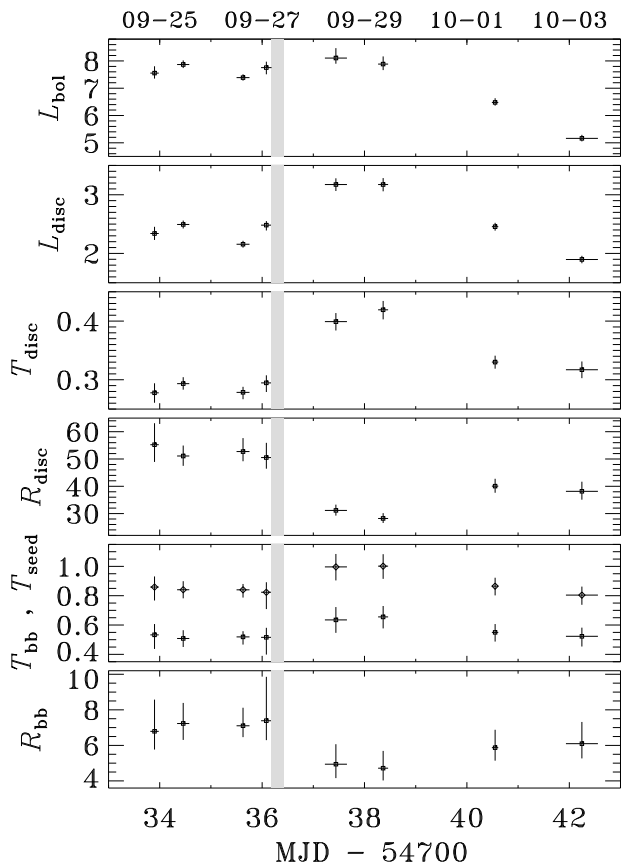}
\caption{Time evolution of the spectral parameters. Luminosities are given in units of $10^{36}\,\ergs$, temperatures in keV and radii in km. The error bars in time axis denote the time between \swift\ and \xte\ pointings. The grey area denotes the time period when the spectral and timing transition must have occurred (see the text and Fig. \ref{fig:ratio_spectra}). The spectral transition stands out as an increase of the inner disc temperature $\Tdisc$, a decrease of $\Rdisc$ and a change in the hotspot radius and temperature. The bolometric luminosity remains roughly constant at $\Lbol \approx 8\times 10^{36} \, \ergs$, which indicates a constant mass accretion rate $\Mdot$ through the transition. Note also that the values of $\Rdisc$ and $\Rbb$ are affected by systematic uncertainties, see text.}
\label{fig:parameters}
\end{figure}

The time evolution of the best-fitting parameters of the spectral modelling are presented in Fig. \ref{fig:parameters} and Table \ref{tab:bestfits}.  
It is immediately obvious that there is a clear transition in the best-fitting spectral parameters on September 27 (MJD 54736), exactly when the timing transition also occurs. 
However, we emphasise that the bolometric luminosity during this transition does not change 
($\Lbol \approx 8\times 10^{36} \, \ergs$ see Fig. \ref{fig:parameters}, top panel). 
The most prominent changes are in the parameters of the accretion disc. 
In the transition, the disc component becomes more luminous, the inner disc temperature $\Tdisc$  increases
and the inner disc radius $\Rdisc$ decreases. 
Before the transition, the bolometric disc luminosity was $\Ldisc \approx 2.4 \times \, 10^{36} \, \ergs$ and after the transition it increased to $\Ldisc \approx 3.2 \times 10^{36} \, \ergs$.
This means that the luminosity of the accretion disc changed from roughly 30 to 40 per cent of the bolometric luminosity, whereas the luminosity of the hotspot decreased.
We also detect a significant increase of the hotspot temperature (and the seed photon temperature) and a decrease in the apparent radius of the hotspot.
We also note that we do not see any change in the optical depth of the shock as $\tau \approx 1$ throughout the outburst (see Table \ref{tab:bestfits}).
This suggests that although the characteristic temperatures and radii change in the transition, the properties in the Comptonized emission above $\sim 10$ keV remain constant.
We make use of this result in Section \ref{sec:pulsemodel}, where we analyse the simultaneous pulse profile transition.

\begin{figure*}
\begin{center}
\begin{tabular}{cc}
\epsfig{file=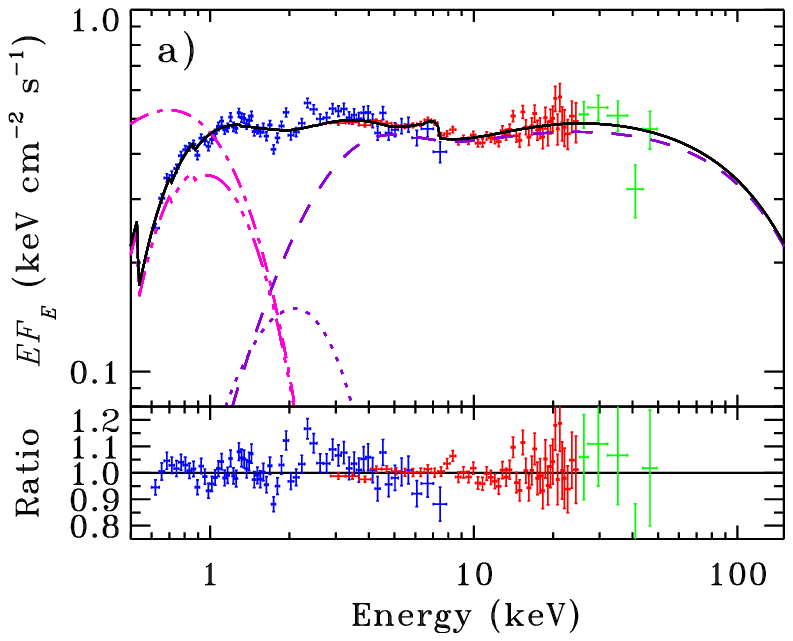,width=0.5\linewidth,clip=} &
\epsfig{file=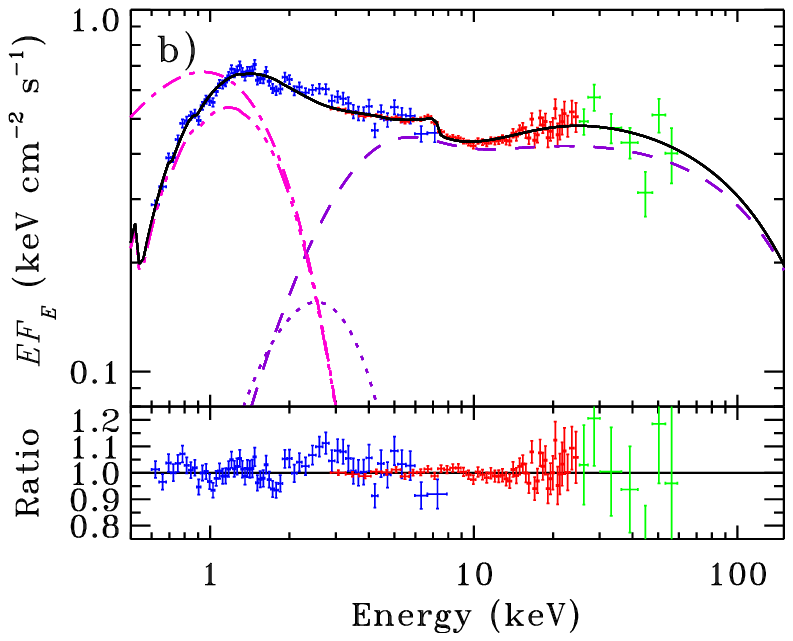,width=0.5\linewidth,clip=} \\
\epsfig{file=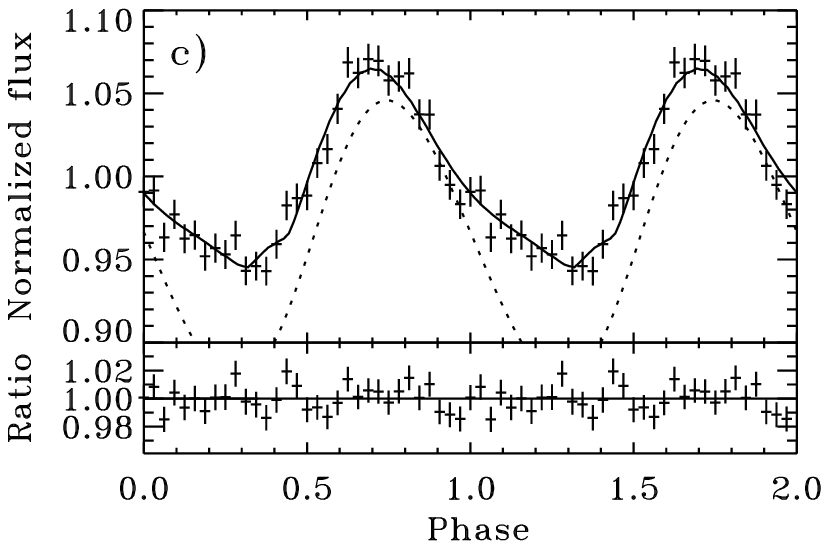,width=0.5\linewidth,clip=} &
\epsfig{file=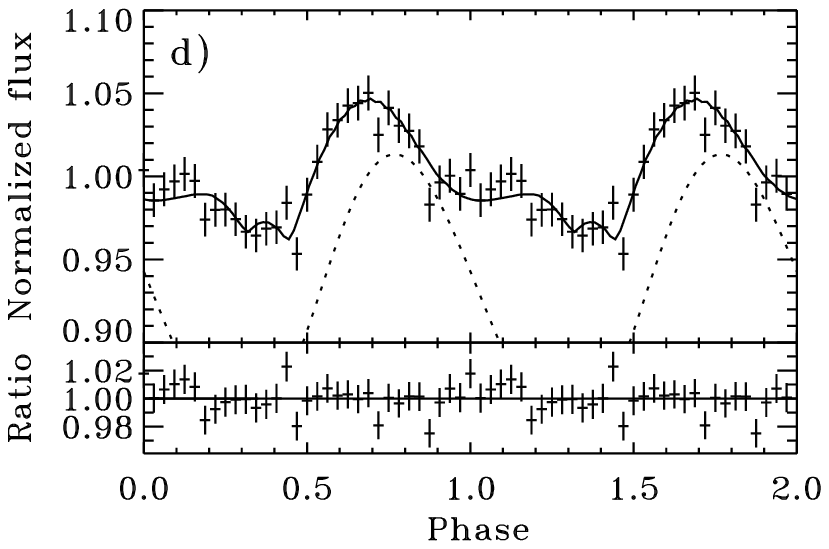,width=0.5\linewidth,clip=} \\
\epsfig{file=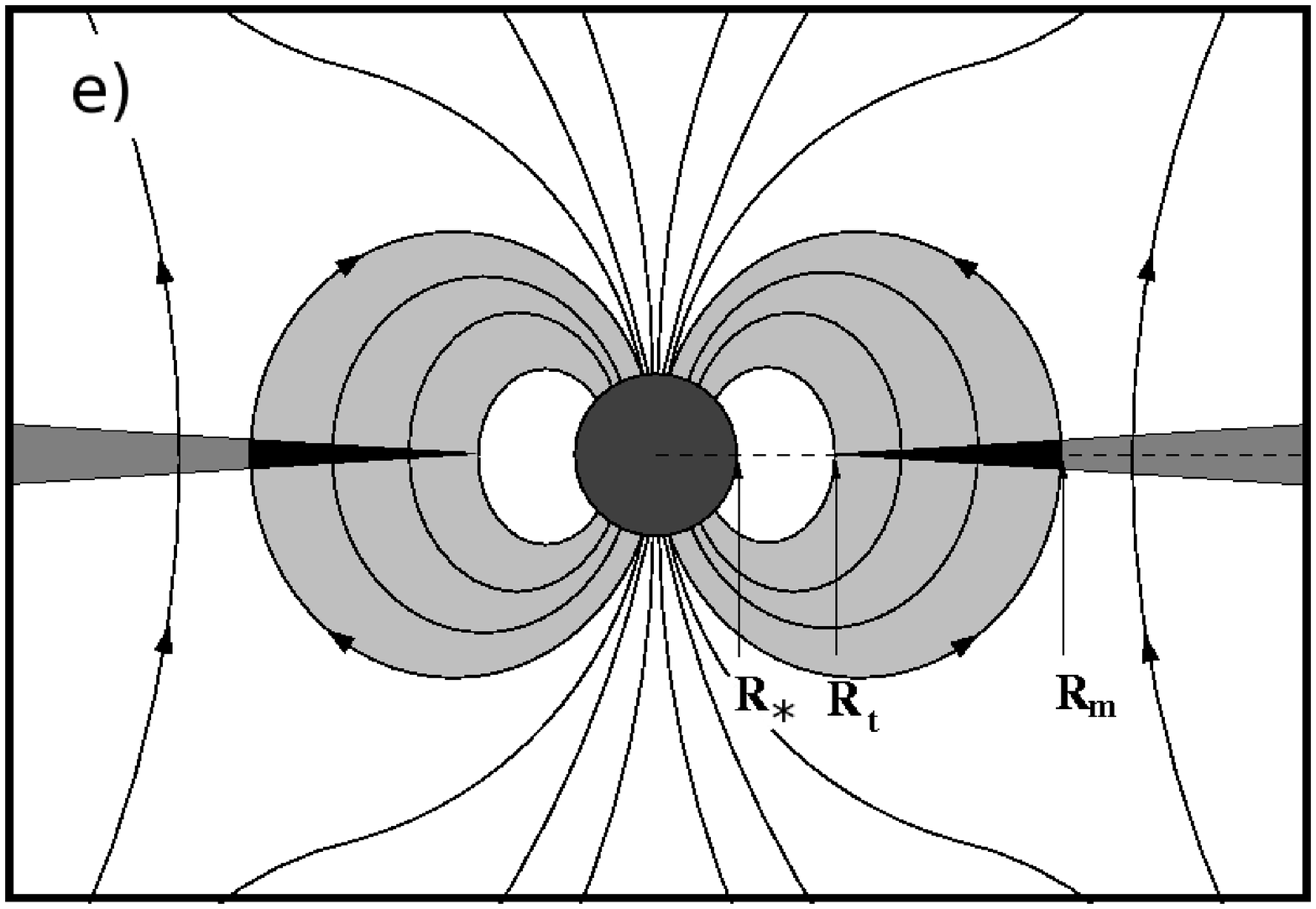,width=0.47\linewidth,clip=} &
\epsfig{file=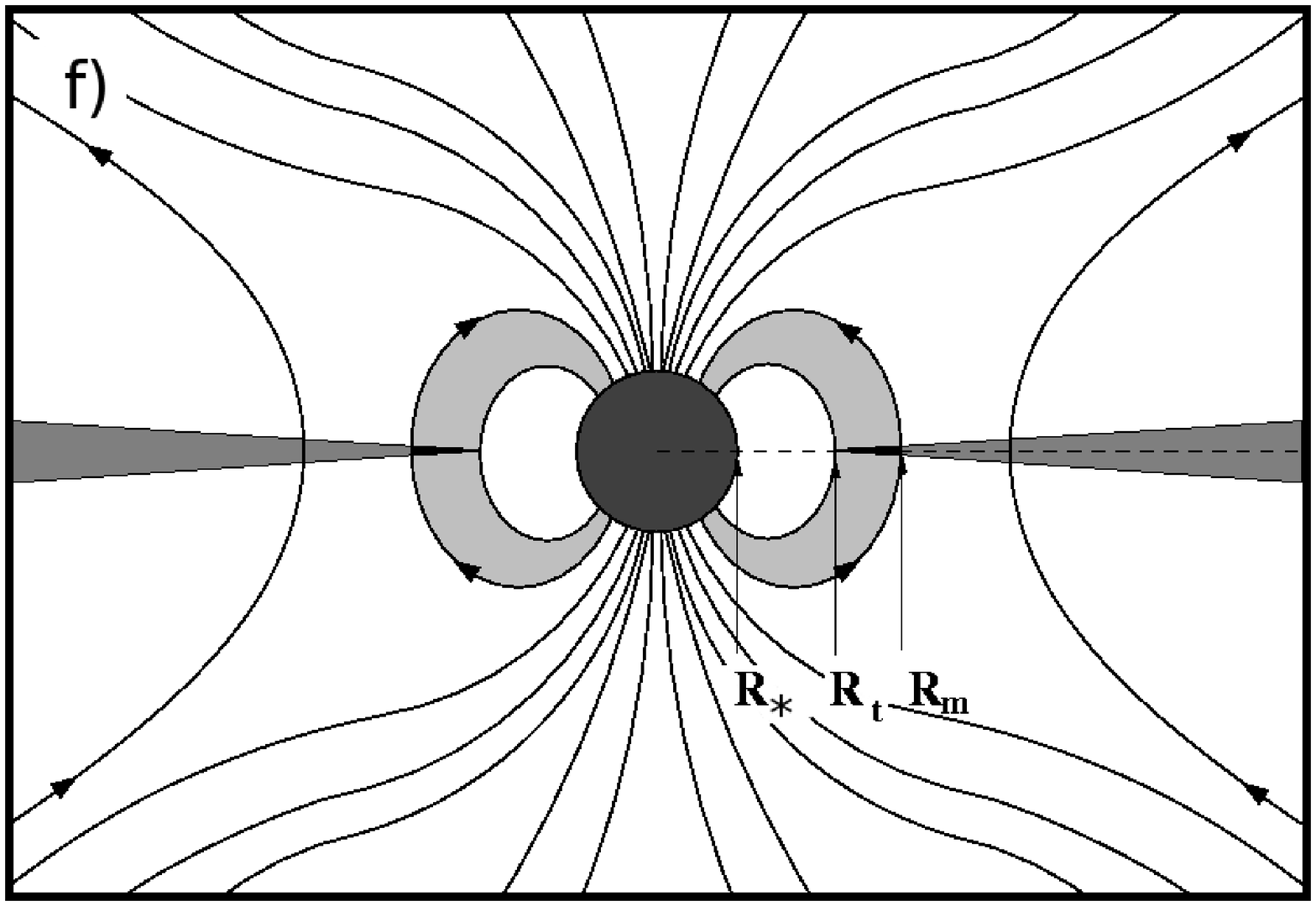,width=0.47\linewidth,clip=}
\end{tabular}
\end{center}
\caption{Photon spectra before the transition (from observations 10 and 11, see Table \ref{tab:obslog}) and after the transition (from observations 13 and 14) are shown in the top panels (a) and (b), respectively. 
The blue, red and green data points are for \swift/XRT, \xte/PCA and \xte/HEXTE, respectively. The solid black line shows the model spectrum. The pink dot-dashed line is the unabsorbed disc component and the three-dot-dashed line the absorbed disc component. The purple dotted line is the NS blackbody component, whereas the dashed line is the Comptonized component. The weak Compton reflection and the iron line are below the plotting range so that the spectral transition is more easily visualized. 
Pulse profile before the transition (from observations 9 and 10) and after the transition (from observation 12) are shown in the middle panels (c) and (d), respectively. 
The solid line is the contribution from both spots, whereas the dotted line shows the contribution of the ``main" spot.  
Lower panels illustrate the geometry of the disc--magnetosphere interaction region before (panel e)  and after (panel f) the transition. 
The dark region in the accretion disc within $\Rt < r < \Rm$ is assumed to rotate as $\omega(r) \approx \omegans$ 
so that the energy dissipation rate in that region is strongly suppressed.}
\label{fig:illustr}
\end{figure*}

The spectral fitting indicates that $\Rdisc$ decreases from $\approx$50 to $\approx$30 km in the transition, while the inner disc temperature $\Tdisc$ increases from $\approx$0.3 to $\approx$0.4 keV.
This effect is seen in both \swift\ and \xte/PCA spectra, which is illustrated in Figures \ref{fig:ratio_spectra}, \ref{fig:parameters} and 
\ref{fig:illustr}. Interestingly, after the transition we see that $\Rdisc \approx \Rco$, as the co-rotation radius 
\be
\Rco = (G\Mns/\omegans^2)^{1/3}
\ee
of \sax\ is about 31 km for a NS mass of $\Mns = 1.4\Msun$ ($\omegans$ is the angular spin frequency of the NS).
There is of course an inherent uncertainty in deriving $\Rdisc$ as its value depends on the assumed distance and the inclination as $\Rdisc \propto D/\sqrt{\cos i}$. 
However, the distance is known rather accurately \citep{GC06} and we will show in Section \ref{sec:pulsemodel} that our initial assumption of $i = 60\degr$ is consistent with the pulse profile modelling. 
Other complications in relating modelled radii (both $\Rdisc$ and $\Rbb$) to ``physical" radii are related to 
the deviation of the local spectra from the blackbody,  assumed boundary conditions and relativistic effects. 
Because these effects do not change the interpretation of the results qualitatively -- in the sense that there is only a systematic shift in the modelled radii -- we defer the discussion of these effects to Section \ref{sec:caveats}.

In accretion on to neutron stars one expects that a constant fraction of the available accretion power is released in photon luminosity; what is not radiated in the disc, will be radiated at the NS surface. 
Therefore, the observed luminosity is thought to be a good measure of the mass accretion rate $\Mdot$. 
As the bolometric luminosity during this transition does not change, we can conclude that the sudden change in the disc parameters cannot be caused by a sudden increase in the mass accretion rate.
This points towards the conclusions that the change in $\Rdisc$ is not related to a change in the truncation radius (which is proportional to the Alfv{\'e}n radius, 
where the pressure of the NS magnetic field equals the ram pressure of the accreting gas, e.g. \citealt{long05}) 
\be \label{eq:alfven}
\Rt = k_{\rm A} (2G\Mns)^{-1/7} \dot{M}^{-2/7} \mu^{4/7}.
\ee
Here $\mu = \Bs \Rns^3$ is the NS magnetic moment and $\Rns$ is the NS radius.
The truncation radius $\Rt$ is also commonly called the magnetospheric radius, but in the following we instead use this term to define the outermost radius $\Rm$ where the field lines of the NS are coupled with the accretion disc \citep[][hereafter LRB95]{LRB95}.
We propose that the interaction region between the disc and the NS magnetic field is large (so that $\Rm > \Rt$, see Fig. \ref{fig:illustr}, bottom panels) and that the modelled inner disc radius actually corresponds to the magnetospheric radius $\Rdisc \sim \Rm$ and not the truncation radius $\Rt$.

\section{Varying disc--magnetosphere coupling in \sax} 
\label{sec:coupling}

\subsection{Physical picture}

We interpret the spectral transition to be caused by a change in the way the dipole field of the NS is coupled with the accretion disc.
The reason for the observed spectral and timing transition is most likely caused by opening of the field lines.

Many theoretical studies of the disc--magnetosphere coupling have been published \citep[e.g.][]{GL79a, GL79b, W87, LRB95,RFS04, KR07}. 
Here we use the framework developed by LRB95 to interpret the observed transition \citep[see][for a review of alternative models]{U04}.
In the LRB95 model, the magnetic field configuration is such that close to the accreting NS the field lines are closed (defined here as the magnetosphere) and outside the magnetosphere the field lines are open (see Fig. \ref{fig:illustr}). 
The field lines decouple from the disc because of a large difference in the angular velocity of the Keplerian disc and the NS (except at $\Rco$).
As the disc is a good conductor, the NS magnetic field is ``frozen" into it and the twisting of the field lines caused by the differential rotation creates a toroidal field component $\Bphi$ out of the poloidal field $\Bz$.
However, there is a critical twist of $\gamma_{\rm c} = \Bphi / \Bz \sim 1$ after which the increased magnetic pressure tends to open the field lines, thus severing the link between the NS and the disc (see LRB95, \citealt{UKL02a, UKL02b, U04} and references therein).

Within the magnetosphere ($r < \Rm$), the magnetic field threading the disc can force the gas to co-rotate with the NS so that $\omega(r) \approx \omegans$ (LRB95), which is also seen in MHD simulations \citep[e.g.][]{RU02}. 
This has a strong effect on the radiation flux emitted from the disc, which is most easily seen from the energy conservation equation (see eq. 7 in LRB95)
\be \label{energy_conservation}
\Sigma \nu_{\rm t} \left( r \frac{\ud \omega}{\ud r} \right)^2 + \frac{4\pi}{c^2} \int_{-H}^H \ud z \, \eta_{\rm t} \bmath{J}^2 \equiv 2 \sigma_{\rm SB} \Teff^4 .
\ee
Here $\Sigma$ is the surface density, $\nu_{\rm t}$ is the turbulent viscosity (which is assumed to be described by the alpha prescription $\nu_{\rm t} = \alpha c_{\rm s} H$, \citealt{SS73}), $c_{\rm s}$ is the sound speed, $z$ is the coordinate along the normal to the disc plane, $H$ is the disc height, $\eta_{\rm t}$ is the magnetic diffusivity of the disc, $\bmath{J}$ is the current density, $\sigma_{\rm SB}$ is the Stefan-Boltzmann constant and $\Teff$ is the effective temperature of the disc. 
In equation (\ref{energy_conservation}) the first term in the left hand side is the viscous dissipation rate per unit area, 
the second term is the Ohmic dissipation rate and the right hand side is the flux radiated by the optically thick and geometrically thin disc. 
If $\omega(r) \approx \omegans$ within the magnetosphere ($r < \Rm$), the viscous dissipation $\propto (\ud \omega / \ud r)^2$ is strongly suppressed and the Ohmic dissipation becomes the dominant energy release mechanism (LRB95).
Expressing the current density through the accretion velocity $v_r$ and vertical magnetic field strength $\Bz = \Bs (\Rns/r)^3$  
as $J=(c^2/4\pi \eta_{\rm t}) (v_r/c) B_z$ and integrating equation (\ref{energy_conservation}) over the radius we get an 
order of magnitude estimate of the luminosity due to the Ohmic dissipation \citep[see also][]{WSL90}:
\be
L_{\rm Ohm} \sim   \frac{H}{\eta_{\rm t}} \int_{\Rt}^{\Rm} v_r^2 B_z^2 r {\rm d}r \sim   
\frac{H v_r^2}{\eta_{\rm t}} \Bs^2 \Rns^2 \frac{\Rns}{\Rt} .
\ee
For geometrically thin disc $H/r \sim 0.01$ with $c_{\rm s} \lesssim 10^8$ cm s$^{-1}$, substituting  
$v_r \sim \alpha c_{\rm s} H/r$ \citep[e.g.][]{FKR02}, $\eta_{\rm t} \sim \nu_{\rm t} = \alpha c_{\rm s} H$ \citep{FS09} and $\alpha <1$, 
we can show that $L_{\rm Ohm} \ll 10^{35}\,\ergs$ for the typical parameters of \sax\ ($\Bs \sim 10^8$ G, $\Rns \approx 10$ km, 
$\Rt \approx 20$ km).\footnote{We constrain the truncation radius to $\Rt \approx 20$ km from the pulse profile modelling in Section \ref{sec:pulsemodel}.} 
Thus the Ohmic dissipation within the magnetosphere is much smaller than the viscous dissipation outside the magnetosphere ($r > \Rm$), 
and therefore does not contribute much to the observed flux from the disc. Therefore, 
the inner disc radius $\Rdisc$ derived from the spectral fits actually should correspond to the magnetospheric radius $\Rm$. 
This makes the physical interpretation of the transition rather straightforward. 

If $\Rdisc \sim \Rm$, the reconfiguration during the transition can be understood as opening of the field lines (see Fig. \ref{fig:illustr}, bottom panels).
Initially the interaction region between the magnetosphere and the disc is large, such that it extends from the truncation radius
$\Rt \approx 20$ km to $\Rm \sim 50$ km.
Then the field lines open resulting in a smaller $\Rm \sim 30$ km, so that the region of the suppressed dissipation 
(where $\omega \approx \omegans$) is smaller (dark regions in the disc in the bottom panels of Fig. \ref{fig:illustr}).
This readily explains why there is an increase in the inner disc temperature $\Tdisc$, a decrease of $\Rdisc$ and why $\Ldisc / \Lbol$ increases.

The fitting indicates that $\Rm\gtrsim\Rco$ before the transition, but we note that the exact value of $\Rm$ is rather uncertain (see Section \ref{sec:caveats}).  
According to \citet{RFS04} $\Rm$ can exceed the corotation radius by 30 per cent before the centrifugal barrier halts the accretion. 
This might set the upper limit for $\Rm$ before the transition. 
Then, the sudden opening of the field lines (decrease of $\Rm$) could be caused by a change in the properties of the accretion disc.
In the LRB95 model $\Rm$ is determined by the critical twist $\gamma_{\rm c} = \Bphi / \Bz = -Hr(\omega(r) - \omegans )/\eta_{\rm t}$. 
As $\gamma_{\rm c} \approx 1$ \citep[e.g.][]{UKL02a, UKL02b} is thought to be a constant, a change in the magnetospheric radius $\Rm$ -- for a fixed $\Mdot$ -- could be caused by a change in the magnetic diffusivity $\eta_{\rm t}$.
Furthermore, as the diffusivity $\eta_{\rm t}$ is most likely caused by the same MHD turbulence that produces the viscosity $\nu_{\rm t}$ (so that $\eta_{\rm t} \sim \nu_{\rm t}$, \citealt{FS09}) we speculate that the observed spectral- and timing transition could be related to changes of viscosity in the accretion disc.
Also, recent MHD simulations have shown that if the magnetic Reynolds number ($= c_{\rm s} H/\eta_{\rm t}$) is below a certain critical value, the magneto-rotational instability (\citealt{BH98}) that is believed to be responsible for the angular momentum transport in the accretion disc might quench, possibly causing high- and low viscosity states \citep{SHB11}.
If the pulse profile variability -- and in general the timing noise -- in AMPs is ultimately caused by such a mechanism, observations of AMPs could be used to place constraints for $\eta_{\rm t}$ and $\nu_{\rm t}$ because for many AMPs the key parameters such as $\Bs$, $\Mdot$ and the relevant radii and temperatures are rather accurately known.

\subsection{Pulse profile modelling} 
\label{sec:pulsemodel}

The observed spectral transition can be caused by a change in the coupling between the magnetic field of the NS and the accretion disc.
Changes in the geometry of the disc--magnetosphere interaction region, on the other hand, alter the way the accretion flow gets channelled onto the NS surface and cause variations in the hotspot size and shape, which affect the pulse profiles (see Fig. \ref{fig:illustr}).
MHD simulations of \citet{RU04} showed that the spot shape strongly depends on the magnetic inclination $\theta$, the angle between the magnetic pole and the rotational axis. 
For $\theta \lesssim 15\degr$ -- which is most likely the case for \sax\ (e.g. \citealt{IP09}) -- the spot shape is close to a ring.

\begin{figure}
\centerline{\epsfig{file=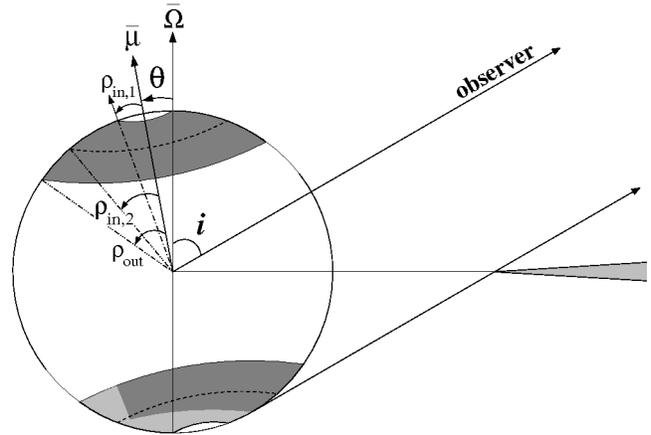,width=\linewidth}}
\caption{Geometry of the hotspots. The magnetic dipole is inclined by angle $\theta$ to the spin axis and 
the co-aligned hotspots are assumed to be ring-shaped. 
The inner edges of the hotspots are displaced from the magnetic poles by angle $\rhoin$ and the outer edges by $\rhout$. 
The accretion disc truncation radius, that determines which part of the secondary hotspot is blocked, is related to $\rhout$ using equation (\ref{xiout}). The pulse profile variations can be modelled simply by changing $\rhoin$ (which is a function of $\Rm$) as shown in Section \ref{sec:pulsemodel}. }
\label{fig:spot_geometry}
\end{figure}

We thus assume that the hotspot shape is a ring around the magnetic pole so that the size of the hotspot is determined by the inner hotspot edge $\rhoin$ and the outer hotspot edge $\rhout$ (see Fig. \ref{fig:spot_geometry}). 
By further assuming that the NS magnetic field is a dipole, $\rhoin$ and $\rhout$ can be related to the magnetospheric radius and 
the disc truncation radius \citep[see e.g.][]{FKR02,PIA09}: 
\be \label{xiout}
\sin \rhoin = \cos \theta \sqrt{\Rns / \Rm}, \quad \sin \rhout = \cos \theta \sqrt{\Rns / \Rt}. 
\ee
We also assume that changes in the magnetosphere size do not affect the truncation radius $\Rt$.
This can be justified as the detailed mechanism of the interaction should not play a big role owing to the strong ($\propto r^{-6}$) radial dependence of the magnetic energy density, which -- for a given $\Mdot$ -- determines the location of $\Rt$.
Obviously parameters such as $\Mns$, $\Rns$, $i$, $\theta$ and $\Bs$ do not change in the transition either. 
What is expected, however, is that the hotspot size should change, which was also seen in the modelling of the time averaged spectra in section \ref{sec:spectra}. This can be seen as a natural consequence of the opening of the field lines.
As illustrated in Fig. \ref{fig:illustr}, the field lines that connect the disc at $\Rm$ are the ones that connect closest to the magnetic pole of the NS.
Given our assumption that $\Rt$ is constant, $\rhout$ is also constant as $\theta$ is constant.
Therefore, under these assumptions, the change in the hotspot size can only be caused by a change of the inner hotspot edge $\rhoin$ (see Figs \ref{fig:illustr} and \ref{fig:spot_geometry}).
Furthermore, the opening of the field lines would not change the hotspot centroid, and thus would not cause a jump in the pulse phases.
As we do not see any such change in phases immediately after the pulse profile change (see Fig. \ref{fig:ampl_phases}), our choice of a hotspot as a symmetric ring around the magnetic pole seems to approximate the emission region to a sufficient accuracy. 

We use the PCA data from observations 9 and 10 for the ``pre-transition" pulse profile and observation 12 for the ``after-transition" pulse profile. 
We only concentrate on the pulse profile changes in the hard band (9.8--23.2 keV) for two reasons.
(1) In the hard band we have emission only from the Comptonized component, whereas in the soft band blackbody component 
also contributes. 
(2) In the hard band we have emission from high scattering orders \citep{VP04,P08}, where the energy and angular distributions of the radiation intensity $I(E,\zeta)$ at the stellar surface can be approximated with a simple formula:   
\be
I(E,\zeta) \propto (1-h \cos \zeta ) E^{-(\Gamma -1)},
\ee
where $E$ is the photon energy, $\zeta$ is the angle relative to the surface normal, $h$ is the anisotropy parameter and $\Gamma$ is the power-law photon index ($\Gamma\approx2$ for \sax).

We fitted the profiles with the model described in detail in \citet{PG03} and \citet{PB06}. 
The model accounts for special relativistic effects (such as Doppler boosting and aberration) as well 
as the general relativistic effects (redshift and light bending). We also account for the eclipses of the hotspot by the accretion 
disc as described by \citet{IP09}. 
We initially arbitrarily chose $\Mns = 1.4 \Msun$, $\Rns = 10.3$ km, $i=60\degr$. 
We set $\theta$ and $\rhout$ as free parameters, but forced them to be the same for the two pulse profiles.
$\Rt$ was computed from $\rhout$ and $\theta$ using equation (\ref{xiout}).
An additional free  parameter is the phase shift $\phi_{\rm shift}$ describing the displacement of the pulse peak. 
Because the extracted pulse profiles were separated by one day and no shifts in phases were seen (see Fig. \ref{fig:ampl_phases}), we force $\phi_{\rm shift}$ to be the same for both pulse profiles.
The only parameters that we initially let to vary in the transition were $\rhoin$ and $h$, which in our view could change if the disc--magnetosphere  coupling changes.

However, the strong secondary pulse after the transition (which we associate with the secondary spot) proved to be problematic to model.
The only way we could get such a peculiar shape was that the secondary spot was partially blocked by the accretion disc (we assume that the secondary spot is antipodal).
We realised that the reason we did not get a good fit to the data was our guess of $\Mns$, $\Rns$ and $i$.
We then let $\Rns$ and $i$ to vary in the fitting and found that there was a rather tight range of these parameters that we could produce such a pulse shape.
We found that for $\Mns=1.4 \Msun$, only with a NS radius of $\Rns \approx 11$ km were we able to produce the pulse shape correctly.
Furthermore, in the initial fitting we did not see a significant change in the $h$ parameter between the two pulses as it varied in a tight range of $h \approx 0.57 \pm 0.05$. 
This is actually expected since $h$ fundamentally depends on the optical depth $\tau$ \citep{P08}, which did not change in the transition (see Fig. \ref{fig:parameters} and Table \ref{tab:bestfits}).
Values of $h \gtrsim 0.5$ correspond to optically thin shocks \citep{P08}, so the values of $\tau \approx 1$ and $h \approx 0.57$, obtained here independently from spectral and pulse profile modelling are consistent with each other.
Therefore, we let $h$ to be free parameter, but required it to be the same for both pulses.

Thus, the modelling was performed with parameters $h$, $\rhout$, $i$, $\theta$ and $\phi_{\rm shift}$ and the parameters that were allowed to vary between the pulses were $\rhoinb$ and $\rhoina$.
The best-fitting parameters of the modelling are presented in Table \ref{tab:pulsefits}.
The corresponding pulse profiles are shown in Figs \ref{fig:illustr}(c) and (d), where the dashed line shows the contribution of the main hotspot, and the solid line shows the sum of the main and the secondary hotspots.
The results indicate that the secondary hotspot is almost entirely blocked (by the NS itself and the accretion disc) around phase 0, and becomes partially visible around phase 0.5. 
The overall contribution of the secondary hotspot to the observed flux increases in the transition from about 5 to 10 per cent
mostly because of the increase of the hotspot temperature (associated with the decreasing size). 
We also see that for the assumed $\Mns=1.4 \Msun$ and $\Rns = 11$ km, we get physically plausible results for all the free parameters.
\begin{enumerate} 
\item The inclination is roughly $i \approx 60\degr$, which is consistent with previous estimations \citep{DHT08,IP09}.
\item The value of the anisotropy parameter $h \approx 0.57$ is also what we expect from Comptonization of optical depth $\tau \approx 1$ \citep{P08}, that was seen in the time averaged spectra.
\item The magnetic inclination $\theta \approx 10\degr$ is consistent with the assumption that the shape of the hotspots are close to a ring \citep{RU04} and similar to previous constraints \citep{IP09}.
\item The truncation radius $\Rt \approx 20$ km, which is within the co-rotation radius of $\Rco = 31$ km and is smaller than $\Rm \approx \Rco$, which is what we expected. 
\item Most importantly, the increase of $\rhoin$ from $< 12\degr$ to $\approx 30\degr$ is what is expected, if the transition in the spectrum and the pulse profile are both caused by a decrease of $\Rm$.
Therefore, we see that the results of the pulse profile modelling are consistent with our interpretation and -- taken together with the spectral information -- fully support the picture that the pulse profile changes in \sax\ are a consequence of a dynamic magnetosphere.
\end{enumerate}

\begin{table} 
 \centering
  \caption{Best-fitting parameters from the pulse profile modelling. }
  \label{tab:pulsefits}
  \begin{tabular}{lll}
  \hline
  Parameter 										& Value & Units \\
  \hline
  NS mass $\Mns$ 									& 1.4 (fixed) 			& $\Msun$ \\ 
  NS radius $\Rns$ 									& 11 (fixed) 			& km \\
  Inclination $i$ 									& $58_{-6}^{+4} $		& deg \\
  Magnetic inclination $\theta$ 					& $11\pm1$ 				& deg \\
  Outer spot radius $\rhout$ 						& $46\pm{6}$ 			& deg \\
  Inner spot radius before  transition $\rhoinb$ 	& $<$12  				& deg \\ 
  Inner spot radius after transition $\rhoina$ 		& $31 \pm 2$ 			& deg \\
  Disc truncation  radius $\Rt$\ $^{a}$ 			& $21_{-3}^{+5}$  		& km \\
  Anisotropy parameter $h$ 							& $0.57 \pm 0.03$  		& \\
  $\chi^2$/d.o.f. 									& 62.7/57				& \\ 
  \hline
  \end{tabular}
  
{$^a$ $\Rt$ is not a free parameter, it was computed using Eq. (\ref{xiout}). }
\end{table}

\section{Discussion}

\subsection{Implications for spin-up torques during the outburst}

The changes in the interaction between the disc and the magnetosphere can have a large effect on the torque that spins up (or spins down) the NS.
Several different models have been developed to compute the torques \citep[e.g.][]{GL79a, GL79b, W87, LRB95, RFS04, MP05, KR07, MP10}.
Generally, in addition to the ``accretion torque" $\tau_{\rm a} = \dot{M}\sqrt{G \Mns \Rt}$ \citep[e.g.][]{MP10}, the disc--magnetosphere  coupling can cause an additional ``magnetic torque," which is strongly dependent on the nature of the interaction and varies significantly between the models.
Nevertheless, it can be concluded that small variations of $\Rm$  independent of the mass accretion rate
(as observed here for \sax) can cause the NS spin-up or spin-down depending on the geometry and position of 
$\Rm$ relative to $\Rco$. 

Also the data suggests, that only those models that predict that $\omega \approx \omegans$ in the magnetosphere can be valid to explain the observed spectral transition.
So observations of AMPs -- as presented here -- can be used to test different hypothesis and constrain the nature of the disc--magnetosphere coupling.

\subsection{Implications for the timing noise in AMPs}

Although the pulse amplitude dropped significantly, the pulse phase remained constant during the transition.
This is actually not surprising, based on the fact that the magnetic inclination seems to be about $\theta \approx 10\degr$.
As the hotspot shape is most likely close to a ring \citep{RU04}, the change in $\Rm$ should only affect the location of the inner edge of the hotspot $\rhoin$, but not change the location of the hotspot on the NS surface. 
However, the situation might be different for other AMPs, where $\theta$ can be larger.
For such AMPs, the hotspot shapes are most certainly not symmetric \citep{RU04}, and changes in $\Rm$ might result in a change in the shape, size, latitude and --most importantly-- the hotspot longitude. Shifts in the hotspot longitude can especially cause large jumps in the observed pulse phases \citep[see][for discussion]{LBW09}.
We speculate that such $\Rm$--dependent motion of the hotspots (and the associated phase jumps) could be the origin of at least part of the timing noise in AMPs.
Observationally, such mechanism could cause the outliers in the X-ray flux -- phase residual relations in XTE J1807--294 \citep{RDB08,PWvdK09,PHW10}, XTE J0929--314 \citep{GC02,PWvdK09} and IGR J17511--3057 (\citealt{RPB11}; \citealt{IKP11}), whereas the overall X-ray flux -- phase residual trends can be caused by $\Mdot$--dependent motion of the hotspots \citep{LBW09,PWvdK09,PHW10}.
Also, we stress that an additional contribution to the timing noise can be caused by $\Mdot$--dependent variation of the truncation radius $\Rt$, as it affects the visibility of the secondary spot \citep{IP09,PIA09}.
These factors do not however exclude the possibility that some AMPs do spin-up during X-ray outbursts (as is most likely the case with IGR J00291+5934, e.g. \citealt{FKP05,BDL07,Pat10b,HGC11,PRB11}).

\subsection{Possible caveats} \label{sec:caveats}

\begin{enumerate}
\item The truncation radius of $\Rt \approx 20$ km was obtained from pulse profile modelling. 
We assumed that this radius is not affected by the change in the disc--magnetosphere coupling that we propose to be the reason for the transition.
We also assumed that the outer edge of the ring-shaped hotspot $\rhout$ is determined from equation (\ref{xiout}), which is not necessarily the case as the field topology should differ from a pure dipole.
These factors can change the constraints on the other parameters.
However, a detailed analysis of these factors should be done by 3D MHD simulations \citep[as in][]{RU04}, but such a study is clearly beyond the scope of the present paper.    

\item As the Ohmic dissipation in the magnetosphere is small, we proposed that $\Rdisc \sim \Rm$. 
However, there are some uncertainties in deriving $\Rdisc$ in addition to those related to spectral modelling in Section \ref{modelling}.
The main systematic uncertainty in estimating $\Rm$ from $\Rdisc$ comes from the fact that $\Tdisc$ is in reality a colour temperature and not the effective temperature of the disc.
Also, the inner disc boundary condition might differ significantly from what is assumed in the {\sc diskbb} model, where the radial dependence of temperature is $T(r) \propto r^{-3/4}$.
These effects could be, in principle, accounted for to obtain the corrected inner disc radius
\be
R'_{\rm disc} =  \xi f_{\rm col}^2 \Rdisc, 
\ee
where  $\xi$ accounts for the fact that the modelled inner disc temperature $\Tdisc$ does not actually occur at $\Rdisc$ and $f_{\rm col}$ is the colour correction factor \citep{G99}.
The colour correction factor was computed to be $f_{\rm col} \approx 1.7$ by \citet{ST95} for black hole systems in the soft state and it seems to be weakly dependent of the mass accretion rate \citep[see][]{DBH05}.
The correction factor $\xi = 0.37$ was computed by \citet{G99} by assuming a zero torque boundary condition at the inner edge of the disc in the pseudo-Newtonian potential.
However, because of the different inner disc boundary conditions for accretion onto an AMP $\xi$
can be much different especially in the beginning of the outburst, where the spectral modelling indicates that $\Rm \gtrsim \Rco$ and therefore $\omega_{\rm K} < \omegans$.
In this case, there should be a region around $\Rm$ where a large jump from the Keplerian rotation to the co-rotation should occur (LRB95, \citealt{LRN10}).
An estimate of the size of this ``transition region" $\Delta r$ was given by LRB95
\be
\Delta r \sim \gamma_{\rm c} \frac{\eta_{\rm t} c_{\rm s}}{\omega_{\rm K} \nu_{\rm t}}.
\ee
For a thin accretion disc $c_{\rm s}  / v_{\rm K}  \sim H / r \ll 1$, so at $r=\Rm$ the transition region $\Delta r \sim H \ll \Rm$ assuming that $\eta_{\rm t}  \sim \nu_{\rm t}$ and $\gamma_{\rm c}  \sim 1$. 
Therefore, we speculate that its effect on the observed radiation flux is small due to the small size, but the exact value of the correction factor $\xi$ is uncertain. 
Furthermore, as the accretion disc is irradiated by the emission of the hotspot and part of the hard emission is absorbed by the disc, the observed colour temperature can be altered and therefore affect $\Rdisc$.
But because the amplitude of Compton reflection $\refl$ is poorly constrained, it is not possible to accurately take this effect into account.
All these unknown factors cause a systematic error in $\Rdisc$, which makes an accurate estimation of $\Rm$ currently infeasible.

Similarly, the hotspot temperature $\Tbb$ is also a colour temperature (at the infinity), but $f_{\rm col}$ is most likely the order of unity in this case \citep[see][and references therein]{GP05} and the actual hotspot size can be larger depending on the stellar compactness and the hotspot geometry \citep[e.g.][]{PG03,GP05,IP09}.
These considerations only change the reported results quantitatively, but the qualitative result of the hotspot size variation shown in Fig. \ref{fig:parameters} is not affected.

\item In our modelling we did not consider any complicated spot shapes. 
Based on the simulations of \citet{RU04} a crescent shaped spot might have been more appropriate to use.
However, this would have required several new parameters to the model, which is not justified based on the quality of the data.
This cannot be improved because the pulse profile changes so rapidly that co-adding more data would not be appropriate.
Also by selecting a broader range in energies than the 9.8--23.2 keV band to improve the statistics cannot be done easily, because the blackbody component (with a different emission pattern) will start contributing from below and there are not enough photons above this range.
Therefore, the only way to address this issue is by making extensive simulations and to extract pulse profiles from all the outbursts of \sax\ and then simultaneously fit them.
\end{enumerate}

\section{Summary}

We have studied spectral and pulse profile variability of \sax\ during its 2008 outburst.
We found that the large drop in the pulse amplitude and the associated change of the pulse profile on September 27 \citep[see fig. 1 in][]{HPC09} was accompanied by a simultaneous spectral transition in the accretion disc, which was most evidently seen in the \swift\ data.
Our interpretation of this transition is that the magnetospheric radius $\Rm$ changes because of opening of the field lines that connect the NS magnetic field to the accretion disc.
We speculated that the physical origin of the field line opening is related to the change of magnetic diffusivity $\eta_{\rm t}$ or viscosity 
$\nu_{\rm t}$. Through this interpretation we could explain why the apparent inner disc radius -- which we associate with the magnetospheric radius $\Rm \sim \Rdisc$ -- decreases from 50 to 30 km in the spectral modelling, why the accretion disc temperature increases from 0.3 to 0.4 keV and why the disc contribution to the bolometric flux increases from 30 to 40 per cent.

We also saw a decrease in the apparent radius of the hotspot in the spectral analysis and this is also consistent with our interpretation.
The field lines that connect the disc at $\Rm$ are the ones that connect closest to the magnetic pole of the NS.
As we assumed that the hotspot shape is a ring around the magnetic pole, the opening of the field lines should change the location of the inner hotspot edge $\rhoin$.
Therefore, as the outer hotspot edge $\rhout$ should not change in the transition -- because for a dipole magnetic field its mostly determined by the mass accretion rate that remained constant -- we can associate the decrease in the hotspot radius to a change in the value of $\rhoin$.
Our pulse profile modelling showed that this is exactly what is required to produce the observed transition.
Furthermore, our results from the pulse profile modelling for the other unknown parameters, such as $\theta$ and $i$, were also consistent with previous estimates.  
We did not try to place constrains on the neutron star mass and radius based on these data, but the identification that varying hotspot size can lead to pulse profile changes has profound implications for determination of these most important parameters.
However, in order to place tight constraints, we should not only concentrate on modelling these two specific pulse profiles, but instead make use of all the outbursts of \sax. 
A detailed re-analysis of these data will the main attention of our future work.

To summarize, for the first time we have found evidence that a sudden pulse profile transition in \sax\ was most likely caused by a change in the way that the NS magnetic field is coupled to the accretion disc.
This mechanism can be one of the causes for pulse profile variability (and the associated timing noise) in other AMPs as well.
We constrained the disc truncation radius at $\Rt = 21_{-3}^{+5}$ km and estimated the magnetospheric radius to be about $\Rm \sim 30$--50 km. 
This would allow in principle to estimate the torques acting on the NS, but the spectroscopic determination of $\Rm$ suffers from several systematic uncertainties that currently prevent such an attempt. 
This, however, could be improved if radial profiles of $\omega$, $\Bz$ and $\Bphi$ in the magnetospheric region ($\Rt < r < \Rm$) were accurately known. 

\section*{Acknowledgements}

This work was supported by the Finnish Graduate School in Astronomy and Space Physics (JJEK), EU FP6 Transfer of Knowledge Project ``Astrophysics of Neutron Stars" MTKD-CT-2006-042722 (AI), V\"ais\"al\"a foundation (MA), and the Academy of Finland grant 127512 (JP). 
AP acknowledges partial support from the Netherlands Organization for Scientific Research (NWO) Veni Fellowship and from a ESF/COMPSTAR visit grant. We thank the referee for helpful comments. 
This research made use of the NASA Astrophysics Data System and of the data obtained from the High Energy Astrophysics Science Archive (HEASARC),  which is a service of the Astrophysics Science Division at NASA/GSFC and the High Energy Astrophysics Division of the Smithsonian Astrophysical Observatory. 

%\bibliographystyle{mn2e}   
%\bibliography{mnemonic,allbib}

\label{lastpage}

\end{document}